# Trust-Based Identity Sharing for Token Grants

1st Kavindu Dodanduwa
Informatics Institute of Technology
Colombo, Sri Lanka
kavindudodanduwa@gmail.com

2nd Ishara Kaluthanthri
9/B Highlevel Road, Godagama
Homagama, Sri Lanka
isharaumadanthi@gmail.com

*Abstract*— Authentication and authorization are two key elements of a software application. In modern day, OAuth 2.0 framework and OpenID Connect protocol are widely adopted standards fulfilling these requirements. These protocols are implemented into authorization servers. It is common to call these authorization servers as identity servers or identity providers since they hold user identity information. Applications registered to an identity provider can use OpenID Connect to retrieve ID token for authentication. Access token obtained along with ID token allows the application to consume OAuth 2.0 protected resources. In this approach, the client application is bound to a single identity provider. If the client needs to consume a protected resource from a different domain, which only accepts tokens of a defined identity provider, then the client must again follow OpenID Connect protocol to obtain new tokens. This requires user identity details to be stored in the second identity provider as well. This paper proposes an extension to OpenID Connect protocol to overcome this issue. It proposes a client-centric mechanism to exchange identity information as token grants against a trusted identity provider. Once a grant is accepted, resulting token response contains an access token, which is good enough to access protected resources from token issuing identity provider's domain.

*Keywords - Identity and Access Management, Authentication, Authorization, OpenID Connect, OAuth 2.0*

## I. INTRODUCTION

In the past, web users used to have multiple user accounts across different websites. To use full privileges, these users went through site-specific user registration process. This registration included defining a username and a password [1]. Site-specific user information maintenance is a burden to users. Users tend to forget credentials or use the same credentials over multiple sites. This created security loop-holes. Development of OAuth, introduction of OpenID was a result of researchers, domain experts trying to solve this problem [1].

OAuth 2.0 framework defines how to delegate access to a third party by a resource owner [2]. If requires, this framework defines the mechanism to be used by a client application to authorize itself to access a protected resource [2]. Separation of authorization layer from client applications enables them to support multiple authorization domains, which support the OAuth framework. Also, with this approach, end user credentials are exposed only to the authorization server. This enables a single user account to be used with multiple services, without the risk of credential exposure to untrusted parties [2] [3].

OpenID Connect extends OAuth 2.0 framework by adding authentication capability [4]. Authentication of OpenID Connect is built on top of ID token [4]. ID token is represented as a JSON Web Token (JWT) [5] and contains details (claims) about the user who authenticate in the OAuth 2.0 request [4]. Since OpenID Connect is built on OAuth 2.0, the token response contains OAuth 2.0 tokens along with ID token. This makes OpenID Connect a protocol that supports both authentication and authorization [4] [6].

Today OAuth 2.0 and OpenID Connect are being widely adopted by various technology platforms. For example, Internet of Things (IoT) is a trending topic in software industry. According to Gartner hype cycle of emerging technologies, as of 2017, IoT is at the peak of inflated expectations [7]. Same source highlights that IoT will be in *plateau of productivity* within next two to five years. IoT will increase the number of connected devices and at the same time, there will be the requirement to provide authentication and authorization for these devices. There are OAuth 2.0 based authorization models suggested for cloud computing [8]. Also, there is research done to identify the performance of OpenID Connect for IoT framework [9] which conclude OpenID Connect as a suitable candidate for IoT platforms. There are researches done to identify the suitability of OAuth 2.0 and OpenID Connect in cloud computing [10], [11]. These researches suggest that OpenID Connect will be a key element in future IoT applications.

As these protocols get widely adopted, it is predictable to see more and more deployments of identity providers supporting these protocols. With the surge of deployment of identity providers, it is foreseeable to see the need of inter-identity provide communication. Requirements will arise to consume services protected by different identity domains. This paper suggests a solution for this by introducing trust-based identity information sharing between identity providers. Sharing of identity details will enable identity provider to issue access token which can be used by applications to consume protected resources. Rest of the paper explains concepts surrounding this idea. It will formally introduce the trust model and present how this can be done on top of OpenID Connect.

## II. BACKGROUND AND PREVIOUS WORK

### A. OAuth 2.0 in brief

OAuth 2.0 is an Internet Engineering Task Force (IETF) standard which is identified by RFC6749 [6]. The initial version (the predecessor) OAuth is identified by RFC5849 [12]. OAuth 2.0 is defined for HTTP based communications and built based on JSON standard. A key highlight of this protocol is the out of the box support for mobile applications. There are few key roles involved in OAuth 2.0 framework. "Table I" introduces and explain these roles

TABLE I. ROLES AND DESCRIPTIONS

| Role | Description |
|---|---|
| Resource owner | An entity with permission to access protected resource/service. Also identified as an end user when this is a human user. |
| Resource server | A server which contains protected resources. Resource server can consume, and grant access based on OAuth 2.0 access tokens. |
| Client | A client is the application which accesses resources from resource server on behalf of the resource owner. It can be a browser-based application, native application (ex: - mobile application) or a server application. |
| Authorization server | The server responsible for authenticating resource owner. It issues OAuth 2.0 tokens for successful authorization requests. |

According to OAuth 2.0 specification, resource server and authorization server can reside in the same server. They can also reside in different domains [6]. Also, it is common to call authorization server with names like identity provider, identity server. Due to this fact, this paper will use these names interchangeably.

The specification defines different authorization grant types to obtain tokens from authorization server [6]. This paper mainly focuses on resource owner involved grant types. Following are the grant types applicable in this category,

- Authorization code grant
- Implicit grant
- Resource owner password credential grant

Successful completion of a grant will result in a token response. A token response contains an access token. Also, depending on grant type, the response could include a refresh token which can be used to renew access tokens when they expire. Once the client obtains an access token, it can use it to access resources from the resource server [6]. The way to use the access token in a HTTP request is defined by RFC6750 [13]. Also, once a protected resource receives an access token, it can use the token introspection endpoint of the authorization server to validate it. This endpoint and validation process is defined by RFC7662 [14].

*B. OpenID Connect in brief*

OpenID Connect brings an authentication layer to OAuth 2.0 through the introduction of JWT based ID token [4]. Client obtains ID token by completing a token obtaining flow defined by the protocol. These flows resemble to grant types defined by OAuth 2.0. An OpenID Connect flow contains additional request parameters such as mandatory *scope* value of *openid* in authorization request [4]. Also, the *response_type* query parameter will be used to differentiate flow type. Following are the flows defined by OpenID Connect protocol,

- Authorization code flow
- Implicit flow
- Hybrid flow

Successful token response from an OpenID Connect flow will contain the ID token. ID token conveys information about token validity as well as authenticated end user at authorization server [4]. Client can validate this ID token against protocol defined steps. The validity of the token will allow the client to authenticate the end user. Claims about end user can be used to desired functionalities such as customizations and welcome messages. Other than the ID token, an OpenID Connect token response can contain an access token. Usage of this access token is similar as if it was obtained through OAuth 2.0 grant type.

*C. Trust between roles*

"Fig.1" highlight the trust relationships already present among OAuth 2 roles.

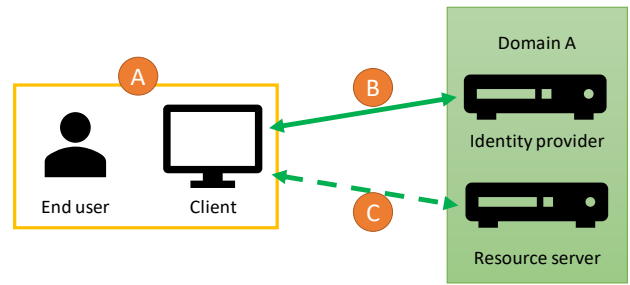

Fig. 1. Trust among OAuth 2.0 roles

As shown in "Fig. 1", the resource owner trusts the client (A). Resource owner lets the client access protected resources on top of this trust relationship. On the other hand, the client and authorization server have a trust relationship (B). The specification defines this to be established through a registration process [6]. Once token obtaining process is done, the client can consume resources from the resources server (C). Trust between resource server and the client is established through OAuth 2.0 tokens. Also, resource server may use other terms to govern this trust through means such as firewalls, VPNs or IP filtering. For example, resource server may only allow requests from a specific IP range to access resources even if the request contains an access token. But these are out of the scope of OAuth 2.0 framework and this paper.

Since OpenID Connect is built on OAuth 2.0, this trust relationship is present in OpenID Connect as well. The addition is the ID token which conveys end user claims (along with claims needed for token validation) which are used to authenticate the end user. ID token helps to build a two-way trust relationship between the resource owner and the client

*D. Previous work*

Most of the existing work in OAuth 2.0 and OpenID Connect focuses on applicability, performance and security concerns of the respective protocols. This is evident from referenced literature such as [2] [3] [8] [9] [10]. There are standards such as SAML which supports the concept of identity federation [15] [16]. Identity fedaration works similar to proposed solution of this paper.

*Trust Requirements in Identity Federation Topologies* [15] describe how identity federation works and how trust models are established in it. As it shows, identity federation allows user identities to accept across different organizational boundaries. This is done through assertions given by trusted parties. Once assertion is accepted, protected resources can be accessed across domains [15] [16]. As shown by the same sources, identity federation concept is present in SAML and WS-Federation, which are standards that existed prior to OAuth 2.0 and OpenID Connect. But the proposed work of this paper is built on top of OpenID Connect which can be considered as a successor to earlier mentioned standards.

*Cloud Federation Management and Beyond: Requirements, Relevant Standards, and Gaps* [16] discuss the impact of identity federation on cloud computing. As it highlights, cloud deployment strategies greatly benefit from identity federation. It highlights the importance of well-governed trust establishment process, monitoring and auditing related to identity federation. This is a critical input for proposed work by this paper.

RFC7521 defines an assertion framework to extend grant types for OAuth 2.0 [17]. This framework defines how assertions from another system can be used by a client to obtain access tokens. Token grant defined by this paper have a similarity with such assertion mechanism. But unlike trying to be a framework, this paper present the trust establishment process, token obtaining process and token grant validation mechanism thus being a complete solution.

III. METHODOLOGY

A. *Involved roles & trust establishment*

There are four primary roles involved in the problem domain. Explaining these roles helps to identify the problem in depth as well as establish how trust establishment should work in the proposed solution. "Table II" introduces these roles and explains their involvement.

TABLE II. PRIMARY ROLES INVOLVED IN THE PROBLEM DOMAIN

| Role | Description |
|---|---|
| Client | An application that represents users. It requires end-user authentication, hence uses OpenID Connect. Also, there is the need for this client to consume services in two domains identified as domain A and domain B |
| Identity provider A | This identity provider governs identities in domain A. It issues OAuth 2.0 tokens that are trusted and accepted by protected services in domain A |
| Identity provider B | This identity provider governs identities in domain B. It issues OAuth 2.0 tokens that are trusted and accepted by protected services in domain B |
| End user | A human user who only have identity details registered in domain A. He/she uses the client and consume resources in domain A. But there is requirement of this user to access protected resources in domain B through the client |

Other than these four primary roles, there will be the involvement of resource serves. They reside inside above mentioned domains and only accepts OAuth 2.0 tokens issued by the domain governing identity provider. As previously mentioned, these resource servers could use token introspection or other valid mechanisms for token validation.

"Fig. 2" shows how primary and secondary roles are involved in a problem scenario.

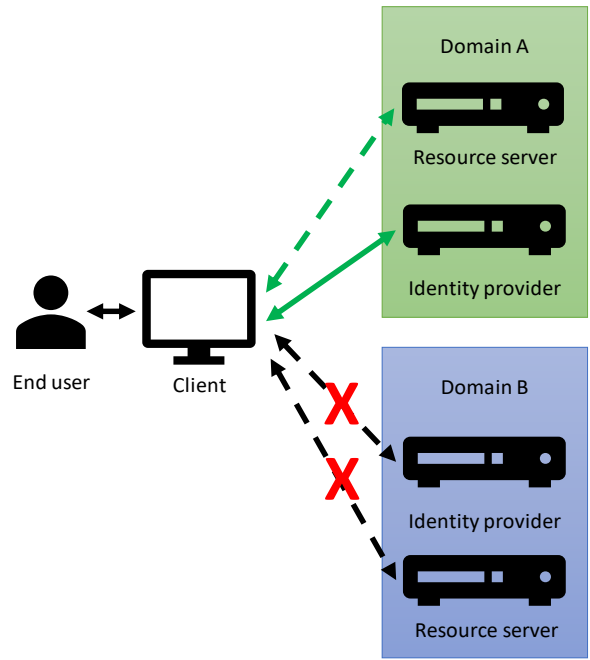

Fig. 2. Primary and secondary roles in a problematic scenario

As highlighted in the "Fig. 2", the end user, the client and roles in domain A have a trust relationship similar to what is shown and explained by "Fig. 1". This allows the client to complete an OpenID Connect flow and obtain access tokens and ID tokens. And it can use obtained access tokens to consume a protected resource from domain A. But there is no trust relationship among client, domain A and domain B. Because of this, client cannot obtain tokens from the identity provider in domain B. Hence it cannot consume protected resources in domain B.

The solution for this is to first establish a trust relationship among primary roles. First, the client must be trusted by identity providers in both domains. This trust relationship allows client to obtain tokens to authenticate against any identity provider. Also, the identity provider in domain A must trust identity provider in domain B. Similarly, identity provider in domain B must trust identity provider in domain A. Trust among identity providers allow them to issue tokens for each other. Once the client obtains such token, client can use that token as token grants against identity provider. This token must contain identity details of the end user. These details are required for monitoring and auditing purposes. Validity of this token grant will issue OAuth 2.0 access token to the client. This access token cab be used to consume resources from the domain. This is the basis for proposed extension by this paper. As previously described, sharing of identity information is essential for the token validation. This justifies the choice of OpenID Connect rather than OAuth 2.0, which focuses only on the authorization.

Establishment of trust could be done in several ways. Trust among client and identity providers can be established through a registration process. Another approach is to use OpenID Connect dynamic client registration protocol [18]. Trust between identity providers can be establish through a registration process. But unlike client registrations, there must be strict policies and validations that must be followed in the process. "Fig 3" formally presents trust relationships in the proposed solution.

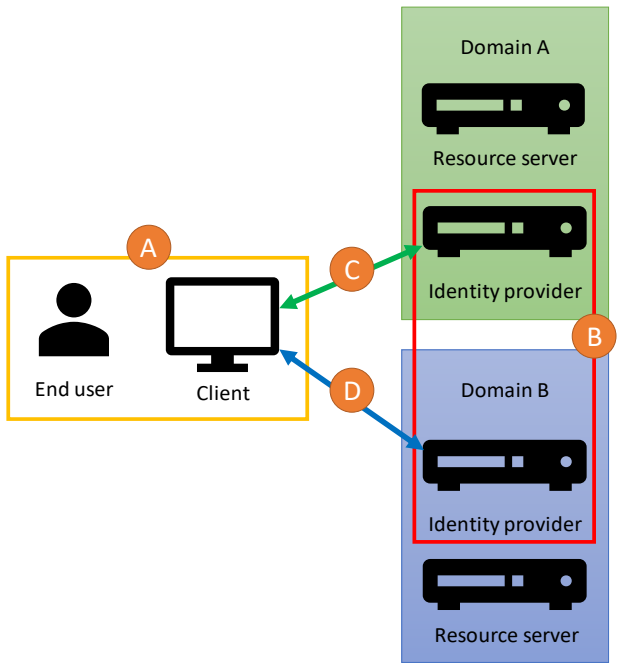

Fig. 3. Trust relationships in the proposed solution

Following list further explains the trust relationships shown in "Fig. 3".

- A – Trust between end user and client. This trust is similar to that exists in OAuth 2.0 and OpenID Connect
- B – Trust between identity providers in domain A and domain B
- C – Trust between client and identity provider of domain A
- D – Trust between client and identity provider of domain B

Also, resource servers will have a trust relationship with the identity provider in the domain. This trust enables clients to consume resource servers with tokens issued by domain governing identity provider.

*B. Identity share token*

Authors propose to use a JWT token to share identity information. JWT is standardized through RFC7519 [5]. Using such established standard benefits adaptability and maintainability of the proposed extension. Furthermore, OpenID Connect already defines user claims that can be included in an ID token. Identity share token can utilize the same user claims defined by OpenID Connect specification. Other than user related claims, identity share token will need to contain claims that are mandatory for token validation and trust establishment. "Table III" contains such claims that authors proposed to have in the token. Adaptations may include other claims that are not mentioned there.

TABLE III MANDATORY CLAIMS IN IDENTITY SHARE TOKEN

| Claim | Description |
|---|---|
| iss | Stands for *issuer*. This claim must contain an identifier unique to the identity share token issuing identity provider |
| aud | Stands for *audience*. This claim contains an identifier unique to the token receiving identity provider. Could be a single value or a JSON array |
| sdata | Stands for *subject data*. This is a JSON object. Authors propose user claims to be included in this JSON object. This claim can be encrypted upon agreements between identity providers. This prevents leakage of information to client application |
| exp | Stands for *expiration*. Contains the token expiration timestamp |
| iat | Stands for *issued at*. Contains the timestamp when the token was created |

Also, the identity sharing JWT token can be signed or encrypted. Choice of this need to be decided at the trust establishment stage. But if client requires token to be inspected, for example to validate the audience, then authors propose to use signed tokens. In such case, *sdata* claim can be encrypted to disallow clients from obtaining sensitive user claims which must only be exposed to token receiving identity provider.

"Fig.4" gives a sample representation of claims included in identity share token.

```
{
    "iss": "https://Domain_A/idp",
    "aud": "https://Domain_B/idp",
    "iat": 1532683271,
    "exp": 1532682999,
    "sdata": {
            "subject": "user1",
            "email" : "sample@sample.com"
        }
}
```

Fig. 4. Identity share token's mandatory claims

*C. Token obtaining process*

Identity share token obtaining is built into usual OpenID Connect flow. To obtain an identity share token, client insert *scope* value *identity_share* in the OpenID connect authentication request. This scope value is mandatory for the proposed extension. Other than that, authorization request can contain an additional query parameter named *identity_share_target* which if present will contain an identifier of the identity provider which identity share token intended to.

Upon receiving the authentication request, the token issuing identity provider must follow all validations defined by OpenID Connect specification. If request contains *identity_share* scope, then identity provider should prepare the identity share token. If query parameter *identity_share_target* is present, the identity provider must verify that the requested target exists in a trusted identity provider list. If the target is present, then *aud* claim of the identity share token must be set to the same target value. "Fig.5" shows an example authentication request targeting token issuing identity provider. It uses authorization code flow.

```
authorization?
   response_type=code
   &client_id=jdf0Plm_op
   &redirect_uri=http%3A%2F%2Fsample.com%2Fredirect%2F
   &scope=openid%20identity_share
   &identity_share_target=http%3A%2F%2Domain_B.com/idp
   &state=pTl987HmQ
   &nonce=12_90oPls
```

Fig. 5 Authentication request with identity share token extension

This token obtaining process can be used with any flow defined by OpenID Connect protocol. Once a flow is completed, the token response must contain the identity share token with a response parameter named *identity_share_token* along with other requested tokens.

*D. Identity share token as token grants*

Once the client obtains identity share token, next step is to use the token as token grants against an identity provider. This grant is defined as *identity share token grant*. Hence it uses a JWT token, it is similar to RFC7523 [19], a profile built on top of assertion framework defined by RFC7521 [17]. RFC7523 defines how JWT tokens can be used as client authentication or authorization grants.

The proposed solution of this paper adopts JWT token validation process from RFC7523 but defines its own token request format.

The token grant request must define *grant_type* as *identity_share_token*. Identity share token is sent with a parameter named *shared_token*. Other than these parameters, the token request must contain parameters relevant to client authentication. Client credentials used for authentication are the ones issued by identity share token consuming identity provider. "Fig.6" shows an example token request with identity share token.

```
POST /token-endpoint HTTP/1.1
Content-Type: application/x-www-form-urlencoded
grant_type=identity_share_token
&shared_token=ertIu87[...omitted for brevity...]
&client_id=8UyfGho2pLqCmNb
&client_secret=uTbC67PqAmbrS1Mx9j2
```

Fig. 6. Token request visualized

Once identity provider receives a token request, it starts the token request validation process. First it validates client_id against known clients. If present and as required by client type, client secrets must be validated. Then identity provider validates grant type to be exactly matched with value *identity_share_token*. If grant type is different, it could be a different grant type. Then request must be validated to contain *shared_token* parameter. If present, content must be identified as a JWT. If the shared token is not present, an error response must be sent with error code *invalid_grant_token*. Then identity provider proceeds to token validation.

Token validation closely follows RFC7523 JWT validation process. First, the identity provider validates *iss* (issuer) claim. Value of this claim must be equal to one of the trusted identity provider. Then *aud* (audience) claim is taken and compared to self-identifier. Claim values *iat* (issued at) and *exp* (expiration) are verified against system clock. Then identity provider must validate JWT token's integrity. As defined by RFC7519, this could be a signature validation, a decryption or both. Next step is to obtain *sdata* (subject data) claim. Depending on trust establishment, value can be encrypted or a plain JSON object. The correct extraction method is detected against validated issuer profile configurations. Once value is obtained, the identity provider will check user claims to contain mandatory claims (ex: - subject, email, age etc.).

The identity provider can solely depend on JWT's integrity to trust user claims present in the token. But further verifications can be done through protocols such as *System for Cross-domain Identity Management* (SCIM) [20]. SCIM protocol defines how to retrieve identity data using known identifiers. This known identifier can the subject claim found in subject data. Such validations further strengthen the trust relationship among identity providers. Once claim validation is done, the identity provider may choose to provision the end user temporarily or permanently for monitoring and auditing purpose.

After a successful token request validation, identity provider will issue a token response containing an access token. This access token should be similar to one obtain through OAuth 2.0 grant or OpenID Connect flow. Token receiving client should be able to consume OAuth 2.0 protected resources using this token. Identity provider may issue a refresh token with the token response to refresh the access token. Token response completes the solution proposed by this paper. "Fig. 7" shows a summary of steps involved in the complete process.

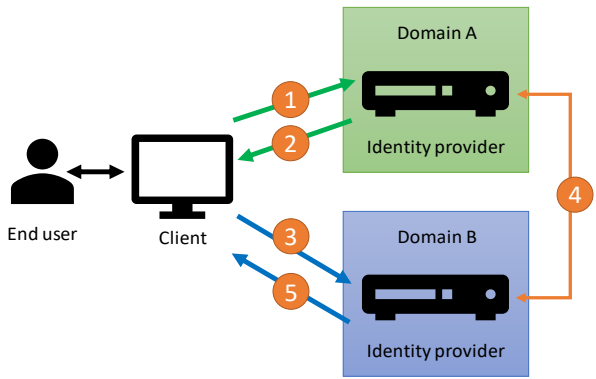

Fig. 7. Summary of steps

1) Client initiates the flow by requesting tokens from domain A. 2) Once the request is validated, identity provider in domain A issue tokens which include identity share token. 3) Client makes a token request to the identity provider in domain B with identity share grant. 4) Identity provider in domain B perform request validations which could include validations against domain B. 5) Identity provider in domain B response to the token request which include an OAuth 2.0 access token.

## IV. CONCLUSION AND FUTURE WORK

This paper proposes a mechanism to obtain OAuth 2.0 tokens through trust-based identity sharing. It defines trust boundaries and proposes trust establishment methods between involved entities. It defines the token format and contents of it. Then it defines the process of obtaining the token, which is built on top of OpenID Connect. Finally, it defines the token request format and token request validation process.

Proposed work should enable client applications to operate against multiple identity providers. This work is mostly based on existing standards such as OAuth 2.0, OpenID Connect, JWT and optionally SCIM. Usage of such standards enhance adoptability of proposed work into existing systems.

As future work, proposed work must be publicly reviewed by domain experts. There will be many improvements that can be done to the proposed solution. For example, security and privacy issues related to proposed mechanism must be identified and rectified.